\documentclass{jpp}
\usepackage{graphicx}

\usepackage[utf8]{inputenc}
\usepackage[T1]{fontenc}
\usepackage{amsmath}

\shorttitle{A generalized Hasegawa-Mima equation in curved magnetic fields}
\shortauthor{N. Sato and M. Yamada}

\title{A generalized Hasegawa-Mima equation\\ in curved magnetic fields}

    \author{Naoki Sato\aff{1}
  \corresp{\email{sato\_naoki@edu.k.u-tokyo.ac.jp}},
  \and Michio Yamada\aff{2}}

\affiliation{
\aff{1}
Graduate School of Frontier Sciences, The University of Tokyo, Kashiwa, Chiba 277-8561, Japan
\aff{2}
Research Institute for Mathematical Sciences, Kyoto University, Kyoto 606-8502, Japan
}

\newcommand{\cp}{\times}

\newcommand{\bol}{\boldsymbol}

\newcommand{\abs}[1]{\left\lvert{#1}\right\rvert}

\newcommand{\lr}[1]{\left({#1}\right)}

\begin{document}

\maketitle

\begin{abstract}
We derive a model equation describing electrostatic plasma turbulence in general (inhomogeneous and curved) magnetic fields 
by analysing the effect of curved geometry on the ion fluid polarization drift velocity. 
The derived nonlinear equation generalizes the Hasegawa-Mima equation governing drift wave turbulence in a straight homogeneous magnetic field, and 
may serve as a toy model for the description of turbulent systems such as the core of H-mode plasmas.
The equation is most appropriate for configurations with a small ExB drift velocity divergence, or a mild spatial change in ExB drift velocity.
We identify the conserved energy of the system, and obtain 
conditions on magnetic field topology for conservation of generalized enstrophy. 
Through numerical examples, we further show how the curvature of the magnetic field reshapes  
self-organized steady turbulent states. 
\end{abstract}

\section{Introduction}
The Hasegawa-Mima equation \citep{HM,HM2} describes $2$-dimensional 
electrostatic plasma turbulence in a straight homogeneous magnetic field. 
Mathematically, it is related to the quasi-geostrophic equation  
for atmospheric dynamics on rotating planetary surfaces \citep{Charney,Charney3}, 
and it reduces to the vorticity equation for a $2$-dimensional incompressible fluid  
in the limit of high electron temperature \citep{Horton}. 
As such, the Hasegawa-Mima equation exhibits properties analogous to
$2$-dimensional fluid turbulence \citep{Batchelor,Kraichnan2,Dritschel}, 
including inverse cascade of energy \citep{Kraichnan,Rivera, Xiao} associated with the presence of two inviscid invariants, 
energy and generalized enstrophy \citep{HM5}. 
Furthermore, in the inviscid limit the system is endowed with a noncanonical Hamiltonian structure \citep{Weinstein,Morrison}, where energy and generalized enstrophy play the roles of Hamiltonian and Casimir invariant respectively \citep{Tassi}. 
These properties combined with a relative simplicity make the Hasegawa-Mima equation an
effective tool in the study of $2$-dimensional fluid and plasma turbulence. Applications include the investigation of turbulence in geophysical flows \citep{Nazarenko} and magnetically confined plasmas \citep{Horton2,Fujisawa}, and more generally the characterization of self-organized turbulent states and zonal flows \citep{Diamond,Singh}. 

Several generalizations of the Hasegawa-Mima equations exist. On one hand, the Hasegawa-Wakatani system \citep{HM4,Wakatani} consists of coupled 
nonlinear equations for the electrostatic potential and the ion density. These equations reveal the interplay between drift wave turbulence and zonal flow mediated by the Kelvin-Helmoholtz instability \citep{Numata}. Hasegawa and Wakatani also included the effect of field curvature in cylindrical geometry to their equations \citep{HM3}. On the other hand, reduced magnetohydrodynamics \citep{Hazeltine} and the four-field model \citep{Hazeltine2} take into account the time-evolution of magnetic flux, parallel ion velocity, and electron pressure. 
Nevertheless, these models are $2$-dimensional, any deviation from a straight magnetic field being treated as a higher-order correction in the relevant ordering. 
Hence, a
closed equation for the electric potential 
describing 
electrostatic plasma turbulence in a general magnetic field is not available at present.   
This deficiency, which
can be ascribed to the rather elusive nature of the polarization drift \citep{Cary,Kauf} in nontrivial magnetic fields, 
has 
made 
the understanding of the impact of topology on the evolution of turbulence a difficult task. Indeed, one usually needs to resort to complete models, such as gyrokinetic theory.  
Furthermore, the development of 
a pertinent toy equation may allow 
a simplified modeling of turbulence in complex plasma systems, such as the core of H-mode plasmas \citep{Bernert} where  density is well approximated by a flat profile.  

The purpose of this paper 
is to fill this gap by deriving an equation describing the evolution of electrostatic turbulence in a general magnetic field
for the simplest plasma system consisting of cold ions and a cloud of electrons obeying the Boltzmann distribution.
As in the classical construction, in principle two approaches could be followed to obtain such equation: guiding center theory or a fluid description. 
In the former, one proceeds by integration of the guiding center distribution function 
to obtain an equation for ion density fluctuations that leads to Hasegawa-Mima type 
electrostatic turbulence by coupling with the gyrokinetic Poisson equation  upon suitable assumptions including cold ions and an adiabatic electron response \citep{Dubin,Krommes}. It should be noted that, in this context, 
turbulence is mediated by a polarization density \citep{Brizard} rather than a polarization drift velocity. 
By contrast, in the fluid approach one starts from the ion continuity equation  
and substitutes the relevant ion $\bol{E}\times\bol{B}$ and polarization drift velocities 
(which are fluid and not guiding center velocities) obtained from ion fluid momentum balance, 
and derives a closed equation for the electrostatic potential (see e.g. \citet{HazeltineBook} 
on the treatment of drift motion within the fluid formalism). 
In this paper we follow the fluid formalism, sticking to the original derivation carried by Hasegawa and Mima (\citet{HM}),  
although we expect an equivalent construction to be possible within the gyrokinetic framework as well.  






 
 \section{Electrostatic potential in curved magnetic fields}
Consider a plasma system consisting of ions and electrons.  
Let $n_e$ denote the electron density. We assume that 
$n_e$ follows a Boltzmann distribution with temperature $T_e$,
\begin{equation}
    n_e=A_e\exp\left\{-\frac{q\phi}{k_BT_e}\right\}.\label{ne}
\end{equation}
Here, $A_e\lr{\bol{x}}$ is a positive real function, $\phi$ denotes the electric potential,  $q=-e$ the electron charge, and $k_B$ the Boltzmann constant. 
It is convenient to introduce the constant
\begin{equation}
    \lambda=\frac{e}{k_BT_e}.\label{lam}
\end{equation}
Let $\bol{B}=\bol{B}\lr{\bol{x}}$ denote a static magnetic field. Recall that $\bol{B}$ must be solenoidal, $\bol{\bol{\nabla}}\cdot\bol{B}=0$.    
In the following, we demand that $\bol{B}\neq0$ throughout the domain of interest $ V$.  
Let $n$ denote the ion density. Assuming quasineutrality, $n_e=Zn$ with $Z$ the number of protons in the ions.
Using \eqref{ne}, the ion continuity equation reads
\begin{equation}
    \lambda\phi_t=-
    \bol{\bol{\nabla}}\cdot\bol{v}-\lambda\bol{v}\cdot\bol{\nabla}\phi-\bol{\nabla}\log{A_e}\cdot\bol{v}
    .\label{ceq}
\end{equation}
Here, $\bol{v}$ denotes the ion fluid velocity.
On the other hand, denoting with $m$ the ion mass, 
the ion fluid equation of motion can be written as
\begin{equation}
m n\frac{d\bol{v}}{dt}=Ze n\lr{\bol{v}\times\bol{B}+\bol{E}}-\bol{\nabla} P,\label{EoMF}
\end{equation}
where $P$ represents pressure and $\bol{E}=-\bol{\nabla}\phi$ the electric field.
It is convenient to decompose the ion fluid velocity as
\begin{equation}
\bol{v}=\bol{v}_{\parallel}+\bol{v}_{\perp},
\end{equation}
where $\bol{v}_{\parallel}$ denotes the velocity component along the magnetic field and $\bol{v}_{\perp}$ the perpendicular one. 
In the drift turbulence setting, the  parallel velocity component is neglected because the time scale of fluctuations $t_d$ is faster than the time scale $t_b$ of ion dynamics along the magnetic field, $t_d<<t_b$.
In the following, 
we will therefore put
\begin{equation}
\bol{v}_{\parallel}=\bol{0},
\end{equation}
and discard the dynamics along $\bol{B}$. 
It is worth noticing that in the opposite regime $t_d>>t_b$ 
the parallel motion $\bol{v}_{\parallel}$ gets averaged out as well due to the
fast bounce oscillation, although the polarization drift becomes neoclassically enhanced due to a large orbit size, leading to a neoclassically modified turbulence model with a dominant Hasegawa-Mima type polarization nonlinearity \citep{Hahm96}.  
In this study we further consider a regime of plasma where the electric potential energy is small compared with the electron kinetic energy, and the ratio between the component of the electric field perpendicular to the magnetic field and the magnetic field itself 
is small,
\begin{equation}
\abs{\frac{e\phi}{k_BT_e}}\sim\epsilon,~~~~\frac{t_d E_{\perp}}{L_{\perp} B}\sim\epsilon,\label{or2}
\end{equation}
where $\epsilon<<1$ is a small ordering parameter, $L_{\perp}$ a characteristic system size across $\bol{B}$, $E_{\perp}=\abs{\bol{E}_{\perp}}$, with $\bol{E}_{\perp}$ the component of the electric field $\bol{E}$ perpendicular to $\bol{B}$, and $B=\abs{\bol{B}}$.
The remaining perpendicular component $\bol{v}_{\perp}$ can be expanded 
into a first-order term $\bol{v}_{1}$ which will correspond to first order $\bol{E}\times\bol{B}$ drift dynamics, a second order term $\bol{v}_{2}$ which will be identified with second order $\bol{E}\times\bol{B}$ drift dynamics plus the polarization drift encountered in guiding center theory, and higher order corrections. at the second order in $\epsilon$ we therefore have  
\begin{equation}
\bol{v}=\bol{v}_{\perp}=\epsilon\bol{v}_{1}+\epsilon^2\bol{v}_{2}.
\end{equation}
To proceed, a further working assumption is needed on the rate of change of $\bol{v}$:
\begin{equation}
\frac{t_d^2}{L_{\perp}}\abs{\frac{\p\bol{v}_1}{\p t}}\sim \epsilon,~~~~\frac{t_d^2}{L_{\perp}}\abs{\frac{\p\bol{v}_2}{\p t}}\sim\epsilon.\label{tv1v2}
\end{equation}
Notice that these conditions are weaker than the scaling adopted in the usual derivation of the Hasegawa-Mima equation, 
$\omega_d/\Omega_c<<1$, with $\omega_d=2\pi/t_d$ and $\Omega_c$ the ion cyclotron frequency. Indeed, $\p\phi/\p t$ is allowed to
contain contributions of order $\epsilon$. This requirement comes from the fact that in a general magnetic field the first order quantity $\nabla\cdot\bol{v}_1$ does not vanish in the continuity equation \eqref{ceq}, and it must therefore be balanced by $\p\phi/\p t$.
The electron fluid will be treated as massless since $m_e<<m$. In particular, we put 
\begin{equation}
m_e=0,
\end{equation}
where $m_e$ is the electron mass. 
Due to the massless assumption for the electron fluid the only contribution to the pressure is that due
to the ions. Furthermore, we enforce the cold ion hypothesis $T=0$, with $T$ the ion temperature, implying that ions are not subject to thermal fluctuations, and therefore 
\begin{equation}
P=0. 
\end{equation}
To obtain expressions for $\bol{v}_{1}$ and $\bol{v}_{2}$, 
consider again the equation of motion \eqref{EoMF}, which at the second order in $\epsilon$ now reads as
\begin{equation}
\epsilon Ze\lr{\bol{B}\times\bol{v}_{1}-\bol{E}_1}=\epsilon^2\left[-m\frac{d\bol{v}_1}{dt}+Ze \lr{\bol{v}_{2}\times\bol{B}+\bol{E}_2}\right].\label{EoMF2}
\end{equation}
Here, we used \eqref{tv1v2} to extract the order of $\p\bol{v}_{1}/\p t$ 
according to $\p\bol{v}_{1}/\p t\rightarrow\epsilon\p\bol{v}_1/\p t$.
In addition, the electric potential $\phi$ and electric field $\bol{E}$ have been decomposed into first order and second order components according to
\begin{equation}
\phi=\epsilon\phi_1+\epsilon^2\phi_2,~~~~\bol{E}=\epsilon\bol{E}_1+\epsilon^2\bol{E}_2=-\epsilon\bol{\nabla}\phi_1-\epsilon^2\bol{\nabla}\phi_2.
\end{equation}
Hence, 
the cross product of equation \eqref{EoMF2} with $\bol{B}/B^2$ can be cast in the form
\begin{equation}
\begin{split}
\epsilon\lr{\frac{\bol{E}_1\times\bol{B}}{B^2}-\bol{v}_{1}}=\epsilon^2\left[-\sigma\frac{\bol{B}\times\frac{d\bol{v}_1}{dt}}{B^2}+\bol{v}_{2}+\frac{\bol{B}\times\bol{E}_2}{B^2}\right],\label{EoMF3}
\end{split}
\end{equation}
where we introduced the physical constant
\begin{equation}
    \sigma=\frac{m}{Ze}.
\end{equation}
Notice that all terms on the left-hand side of equation \eqref{EoMF3} are first order in $\epsilon$, while those on the right-hand side are second order. 
The first order $\bol{E}\times\bol{B}$ flow $\bol{v}_{\bol{E}_1}$ can be obtained from the left-hand side: 
\begin{equation}
\bol{v}_{1}=\bol{v}_{\bol{E}_1}=\frac{\bol{E}_1\times\bol{B}}{B^2}.\label{vE1}
\end{equation}
Similarly, the second order fluid drift is given by
\begin{equation}
\bol{v}_{2}=\bol{v}_{\rm pol}+\bol{v}_{\bol{E}_2}=\sigma\frac{\bol{B}\times\frac{d\bol{v}_1}{dt}}{B^2}+\frac{\bol{E}_2\times\bol{B}}{B^2}.\label{vpol}
\end{equation}
Evidently, the second term on the right-hand side is a second order $\bol{E}\times\bol{B}$ drift $\bol{v}_{\bol{E}_2}$, while
the first term is nothing but  
a 
polarization drift $\bol{v}_{\rm pol}$. Indeed, in a straight magnetic field $\bol{B}=B_0\bol{\nabla} z$ with $B_0\in\mathbb{R}$ one has  
\begin{equation}
\bol{v}_{\rm pol}=\sigma\frac{\bol{B}\times\frac{d\bol{v}_1}{dt}}{B^2}=\frac{\sigma}{B_0^2}\frac{d\bol{E}_{1\perp}}{dt},
\end{equation}
with $\bol{E}_{1\perp}$ the component of $\bol{E}_1$ perpendicular to $\bol{B}$. 
Combining equations \eqref{vE1} and \eqref{vpol}, 
the total ion velocity has expression
\begin{equation}
\begin{split}
\bol{v}&=\epsilon\bol{v}_{\bol{E}_1}-\epsilon^2\sigma\frac{\bol{B}\cp\left[\bol{v}_{\bol{E}_1}\cp\lr{\bol{\nabla}\cp\bol{v}_{\bol{E}_1}}\right]}{B^2}-\epsilon^2\sigma\frac{\p}{\p t}\frac{\bol{\nabla}_{\perp}\phi_1}{B^2}+\epsilon^2\sigma\frac{\bol{B}\times\bol{\nabla}\lr{\bol{v}_{\bol{E}_1}^2+\phi_2}}{2B^2}\\
&=\epsilon\lr{1-\epsilon\sigma\frac{\bol{B}\cdot\bol{\nabla}\cp\bol{v}_{\bol{E}_1}}{B^2}}\bol{v}_{\bol{E}_1}-\epsilon^2\sigma\frac{\p}{\p t}\frac{\bol{\nabla}_{\perp}\phi_1}{B^2}+\epsilon^2\sigma\frac{\bol{B}\times\bol{\nabla}\lr{\bol{v}_{\bol{E}_1}^2+\phi_2}}{2B^2},\label{totv0}
\end{split}
\end{equation}
where we 
introduced 
the orthogonal gradient operator
\begin{equation}
\bol{\nabla}_{\perp}f=\frac{\bol{B}\cp\lr{\bol{\nabla} f\cp\bol{B}}}{B^2},
\end{equation}
with $f$ some function.
Then, we have
\begin{equation}
\begin{split}
    \bol{\nabla}\cdot\bol{v}=&
    \bol{\nabla}\cdot\left[
    \epsilon\lr{1-\epsilon\sigma\frac{\bol{B}\cdot\bol{\nabla}\cp\bol{v}_{\bol{E}_1}}{B^2}}\bol{v}_{\bol{E}_1}
    \right]
    -\epsilon^2\sigma\frac{\p}{\p t}\bol{\nabla}\cdot\lr{\frac{\bol{\nabla}_{\perp}\phi_1}{B^2}}\\&+
    \frac{1}{2}\epsilon^2\sigma\bol{\nabla}\lr{\bol{v}_{\bol{E}_1}^2+\phi_2}\cdot\bol{\nabla}\times\lr{\frac{\bol{B}}{B^2}},
    \label{divv0}
    \end{split}
\end{equation}
Defining the total $\bol{E}\times\bol{B}$ drift velocity as
\begin{equation}
\bol{v}_{\bol{E}}=\epsilon\bol{v}_{\bol{E}_1}+\epsilon^2\bol{v}_{\bol{E}_2}=\frac{\bol{E}\times\bol{B}}{B^2},
\end{equation}
at the second order equation \eqref{totv0} can be expressed solely in terms of $\phi$ as
\begin{equation}
\begin{split}
\bol{v}&=\bol{v}_{\bol{E}}-\sigma\frac{\bol{B}\cp\left[\bol{v}_{\bol{E}}\cp\lr{\bol{\nabla}\cp\bol{v}_{\bol{E}}}\right]}{B^2}-\sigma\frac{\p}{\p t}\frac{\bol{\nabla}_{\perp}\phi}{B^2}+\sigma\frac{\bol{B}\times\bol{\nabla}\bol{v}_{\bol{E}}^2}{2B^2}\\
&=\lr{1-\sigma\frac{\bol{B}\cdot\bol{\nabla}\cp\bol{v}_{\bol{E}}}{B^2}}\bol{v}_{\bol{E}}-\sigma\frac{\p}{\p t}\frac{\bol{\nabla}_{\perp}\phi}{B^2}+\sigma\frac{\bol{B}\times\bol{\nabla}\bol{v}_{\bol{E}}^2}{2B^2},\label{totv}
\end{split}
\end{equation}
We remark that this same result can also be obtained by a simple iteration method
where, regarding the left-hand side of \eqref{EoMF} as a smaller term than the right-hand side, we have the lowest
order solution as $\bol{v}=\bol{v}_{\bol{E}}$ and the next order solution
as given by \eqref{totv}.
Similarly, in terms of $\phi$ equation \eqref{divv0} can be written as
\begin{equation}
\begin{split}
    \bol{\nabla}\cdot\bol{v}=&
    \bol{\nabla}\cdot\left[
    \lr{1-\sigma\frac{\bol{B}\cdot\bol{\nabla}\cp\bol{v}_{\bol{E}}}{B^2}}\bol{v}_{\bol{E}}
    \right]
    -\sigma\frac{\p}{\p t}\bol{\nabla}\cdot\lr{\frac{\bol{\nabla}_{\perp}\phi}{B^2}}\\&+
    \frac{1}{2}\sigma\bol{\nabla}\bol{v}_{\bol{E}}^2\cdot\bol{\nabla}\times\lr{\frac{\bol{B}}{B^2}}.
   \label{divv}
    \end{split}
\end{equation}
On the other hand, the term
\begin{equation}
    \lambda\bol{v}\cdot\bol{\bol{\nabla}}\phi
    =\epsilon^3\lambda\lr{\bol{v}_1\cdot\bol{\nabla}\phi_2+\bol{v}_2\cdot\bol{\nabla}\phi_1+\epsilon\bol{v}_2\cdot\bol{\nabla}\phi_2},\label{vdphi}
\end{equation}
appearing on the right-hand side of \eqref{ceq} is a third-order contribution that can be neglected.
Using equations \eqref{divv} and \eqref{vdphi} to evaluate \eqref{ceq}, at the second order we thus obtain
\begin{equation}
\begin{split}
    \frac{\p}{\p t}\left[\lambda A_e\phi
    -\sigma\bol{\nabla}\cdot\lr{\frac{A_e\bol{\nabla}_{\perp}\phi}{B^2}}\right]=&
    \bol{\nabla}\cdot\left[
    A_e\lr{\sigma\frac{\bol{B}\cdot\bol{\nabla}\cp\bol{v}_{\bol{E}}}{B^2}-1}\bol{v}_{\bol{E}}
    -\sigma A_e\frac{\bol{B}\times\bol{\nabla}\bol{v}_{\bol{E}}^2}{2B^2}
    \right].
    \label{DWT1}
\end{split}
\end{equation}
To check the consistency of the ordering assumptions leading to \eqref{DWT1} we must verify that   
the ion fluid conservation laws are satisfied by the derived equation. 
First consider the total mass
\begin{equation}
M_V=m\int_{V}n\,dV.
\end{equation}
In this notation  $V\subset\mathbb{R}^3$ is a bounded domain occupied by the plasma and ${\rm d}V$ the volume element in $\mathbb{R}^3$.
The constantcy of $M$ follows immediately by noting that at leading order in $\phi$  
\begin{equation}
M_V=m\int_{V}A_e\lr{1+\lambda\phi}\,dV.\label{MV}
\end{equation}
Then, using \eqref{DWT1} and \eqref{totv} we have
\begin{equation}
\frac{dM_V}{dt}=-m\int_{\p V}A_e\bol{v}\cdot\bol{n}\,dS,
\end{equation}
where $\bol{n}$ is the unit outward normal to the boundary $\p V$, ${\rm d}S$ the surface element on $\p V$, and $\bol{v}$ is given by \eqref{totv}. 
This boundary integral vanishes if the system is periodic or   
the fluid velocity $\bol{v}$ is tangential to the bounding surface, $\bol{v}\cdot\bol{n}=0$ on $\p V$, or $A_e=0$ on $\p V$. 
Next, consider the fluid energy
\begin{equation}
E_V=\int_{V}n\lr{\frac{1}{2}m\bol{v}^2+Ze\phi}\,dV.
\end{equation}
To study conservation of energy it is convenient to subtract to $E_V$ the total mass $M_V$ and define a new quantity
\begin{equation}
H_V=\frac{E_V}{Ze}-\frac{M_V}{m\lambda}=\int_V n\lr{\frac{1}{2}\sigma\lambda\bol{v}^2+\lambda\phi^2-\frac{1}{\lambda}}\,dV.
\end{equation}
At leading order in $\phi$ we therefore obtain
\begin{equation}
H_{V}=\frac{1}{2}\int_{V}A_e\lr{\sigma\bol{v}_{\bol{E}}^2+\lambda\phi^2-\frac{2}{\lambda}}\,dV.
\end{equation}
Since $A_e\lr{\bol{x}}$ is independent of time, the quantity $H_V$ can be further simplified to
\begin{equation}
H_{V}=\frac{1}{2}\int_{V}A_e\lr{\lambda\phi^2+\sigma\frac{\abs{\bol{\nabla}_{\perp}\phi}^2}{B^2}}\,dV.\label{HV}
\end{equation}
Using \eqref{DWT1}, the rate of change in $H_V$ is
\begin{equation}
\begin{split}
\frac{dH_V}{dt}=&\int_{V}A_e\lr{\lambda\phi\phi_t+\frac{\sigma}{B^2}\bol{\nabla}\phi\cdot\bol{\nabla}_{\perp}\phi_t}\,dV\\
=&\int_{V}{\phi\left[\lambda A_e\phi_t-\sigma\bol{\nabla}\cdot\lr{\frac{A_e\bol{\nabla}_{\perp}\phi_t}{B^2}}\right]}\,dV+\sigma\int_{\p V}A_e\phi\frac{\bol{\nabla}_{\perp}\phi_t}{B^2}\cdot\bol{n}\,dS\\
=&\int_{V}\phi\bol{\nabla}\cdot\left[A_e\lr{\sigma\frac{\bol{B}\cdot\bol{\nabla}\times\bol{v}_{\bol{E}}}{B^2}-1}\bol{v}_{\bol{E}}-\sigma A_e\frac{\bol{B}\times\bol{\nabla}\bol{v}^2_{\bol{E}}}{2B^2}\right]\,dV\\&+\sigma\int_{\p V}A_e\phi \frac{\bol{\nabla}_{\perp}\phi_t}{B^2}\cdot\bol{n}\,dS\\
=&\int_{\p V}A_e\phi\left[\lr{\sigma\frac{\bol{B}\cdot\bol{\nabla}\times\bol{v}_{\bol{E}}}{B^2}-1}\bol{v}_{\bol{E}}-\sigma\frac{\bol{B}\times\bol{\nabla\bol{v}_{\bol{E}}^2}}{2B^2}+\sigma\frac{\bol{\nabla}_{\perp}\phi_t}{B^2}\right]\cdot\bol{n}\,dS\\
&-\frac{1}{2}\sigma\int_{V}A_e\bol{v}_{\bol{E}}\cdot\bol{\nabla}\bol{v}_{\bol{E}}^2\,dV\\
=&-\int_{\p V}A_e\phi\,\bol{v}\cdot\bol{n}\,dS-\frac{1}{2}\sigma\int_{V}A_e\bol{v}_{\bol{E}}\cdot\bol{\nabla}\bol{v}_{\bol{E}}^2\,dV.
\end{split}
\end{equation}
Here, $\bol{v}$ is given by \eqref{totv} and the notation $\phi_t=\p\phi/\p t$ has been used. 
Notice that boundary integrals vanish if the system is periodic or $\bol{v}\cdot\bol{n}=0$ on $\p V$ or $A_e\phi=0$ on $\p V$.

However, the last term on the right-hand side cannot be written as a boundary integral. Hence, 
an additional ordering condition is needed to ensure that energy is preserved. 
This term originates from the quantity
\begin{equation}
\sigma\bol{\nabla}\cdot\lr{A_e\frac{\bol{B}\times\bol{\nabla}\bol{v}_{\bol{E}}^2}{2B^2}}=\frac{1}{2}\sigma\bol{\nabla}\bol{v}_{\bol{E}}^2\cdot\bol{\nabla}\times\lr{\frac{A_e\bol{B}}{B^2}},
\end{equation}
appearing on the right-hand side of \eqref{DWT1}. We therefore demand that
\begin{equation}
\frac{t_d\sigma}{2A_e}\abs{\bol{\nabla}\bol{v}_{\bol{E}}^2\cdot\bol{\nabla}\times\lr{\frac{A_e\bol{B}}{B^2}}}\sim\epsilon^3.\label{orX}
\end{equation}
Observe that the scaling above reflects the degree of accuracy of the ordering assumption $\p\bol{v}_{2}/\p t\sim\epsilon$ adopted in equation \eqref{tv1v2} which led to the omission of this term from 
fluid momentum balance.
Recalling that $\bol{v}_{\bol{E}}^2$ is second order in $\epsilon$, if the density gradient is small $\abs{\bol{\nabla}_{\perp}A_e}/A_e\sim\epsilon$ the condition \eqref{orX} is automatically satisfied for those magnetic fields such that 
$\abs{\bol{\nabla}\times\lr{B^{-2}\bol{B}}}\sim\epsilon$, or equivalently
\begin{equation}
\bol{B}=B^2\lr{\bol{\nabla}\zeta+\epsilon\bol{\xi}}\label{orX1}
\end{equation}
for some potential $\zeta$ and vector field $\bol{\xi}$. 
This condition implies that the quantity $\bol{B}/B^2$ does not depart largely from a potential field, 
physically implying that the divergence of the $\bol{E}\times\bol{B}$ velocity scales as $\bol{\nabla}\cdot\bol{v}_{\bol{E}}=\epsilon\bol{\nabla}\times\bol{\xi}\cdot\bol{\nabla}\phi\sim\epsilon^2$. 
This is the case of the standard Hasegawa-Mima equation where $\abs{\bol{\nabla}_{\perp} A_e}/A_e\sim\epsilon$ and \eqref{orX} is satisfied with  $\zeta=B_0^{-1}z$, $B_0\in\mathbb{R}$, and $\bol{\xi}=\bol{0}$ so that $\bol{\nabla}\cdot\bol{v}_{\bol{E}}=0$. 
An alternative way to fulfill \eqref{orX} is to assume that
\begin{equation}
L_{\perp}\abs{\bol{\nabla}_{\perp}\bol{v}_{\bol{E}}^2}\sim\epsilon\bol{v}_{\bol{E}}^2,\label{orX2}
\end{equation}
implying a mild change in $\bol{E}\times\bol{B}$ drift velocity across $\bol{B}$.

Once the prescription \eqref{orX} is enforced, the last term in \eqref{DWT1} can be neglected since it scales as $\epsilon^3$, and we obtain a consistent equation for the potential $\phi$,
\begin{equation}
\frac{\p}{\p t}\left[\lambda A_e\phi
    -\sigma\bol{\nabla}\cdot\lr{\frac{A_e\bol{\nabla}_{\perp}\phi}{B^2}}\right]=
    \bol{\nabla}\cdot\left[
    A_e\lr{\sigma\frac{\bol{B}\cdot\bol{\nabla}\cp\bol{v}_{\bol{E}}}{B^2}-1}\bol{v}_{\bol{E}}
    \right],
    \label{DWT2}
\end{equation}
which preserves the total mass \eqref{MV} and the energy \eqref{HV} exactly under suitable boundary conditions. In particular, one obtains
\begin{equation}
\frac{dM_{V}}{dt}=-m\int_{\p V}A_e\bol{v}'\cdot\bol{n}\,dS,~~~~\frac{dH_{V}}{dt}=-\int_{\p V}A_e\phi\,\bol{v}'\cdot\bol{n}\,dS,
\end{equation}
with the effective ion fluid velocity
\begin{equation}
\bol{v}'=\lr{1-\sigma\frac{\bol{B}\cdot\bol{\nabla}\cp\bol{v}_{\bol{E}}}{B^2}}\bol{v}_{\bol{E}}-\sigma\frac{\p}{\p t}\frac{\bol{\nabla}_{\perp}\phi}{B^2}.\label{totv2}
\end{equation}
The conditions on magnetic field topology for existence of an additional conserved quantity (generalized enstrophy) will be discussed in section 4. 

We suggest that equation \eqref{DWT2} is appropriate to describe electrostatic turbulence in magnetic fields with arbitrary topology. In particular, considering the ordering condition \eqref{orX}, the equation is best suited for magnetic fields of the type \eqref{orX1} so that $\bol{\nabla}\cdot\bol{v}_{\bol{E}}\sim\epsilon^2$, or for configurations \eqref{orX2} such that the spatial change in $\bol{E}\times\bol{B}$ drift velocity scales as $\bol{\nabla}_{\perp}\bol{v}_{\bol{E}}^2\sim\epsilon^3$.   

Regarding the physical interpretation of the terms appearing in the equation, 
it is worth observing that the quantity $\bol{\nabla}\cdot\lr{B^{-2}\bol{\nabla}_{\perp}\phi}$   
arises from the component of the vorticity $\bol{\nabla}\times\bol{v}_{\bol{E}}$ along $\bol{B}$. Indeed, 
\begin{equation}
\frac{\bol{B}}{B^2}\cdot\bol{\nabla}\times\bol{v}_{\bol{E}}=\bol{\nabla}\cdot\lr{\frac{\bol{\nabla}_{\perp}\phi}{B^2}}+\left[\bol{\nabla}\times\lr{\frac{\bol{B}}{B^2}}\right]\times\frac{\bol{B}}{B^2}\cdot\bol{\nabla}_{\perp}\phi.
\end{equation}
In addition, a term $\bol{\nabla}\cdot\lr{nB^{-2}\bol{\nabla}_{\perp}\phi}$ is often encountered in modern gyrokinetic and gyrofluid theories as the ion polarization charge density in Poisson's equation for the electric potential \citep{Hahm09,Strinzi}. 
The factor $1-\sigma B^{-2} \bol{B}\cdot\bol{\nabla}\times\bol{v}_{\bol{E}}$ on the right-hand side of \eqref{DWT2} is similarly predicted by gyrokinetic theory as a correction caused by the polarization charge to the magnetic field strength, 
\begin{equation}
B_{\parallel}^{\ast}\simeq B\lr{1+\sigma \frac{\bol{B}\cdot\bol{\nabla}\times\bol{v}_{\bol{E}}}{B^2}}\bol{v}_{\bol{E}},
\end{equation}
which corresponds to the Jacobian determinant of the guiding center phase space when parallel dynamics is ignored. This correction causes a net drift velocity (see equation (4) of \citet{Hahm962} or the discussion in  \citet{Little})
\begin{equation}
\tilde{\bol{v}}_{\bol{E}}=\frac{\bol{B}\times\bol{\nabla}\phi}{BB_{\parallel}^{\ast }}\simeq \lr{1-\sigma \frac{\bol{B}\cdot\bol{\nabla}\times\bol{v}_{\bol{E}}}{B^2}}\bol{v}_{\bol{E}}.
\end{equation}

\section{Limit to the Hasegawa-Mima equation and curvature effects}
In this section we will show that equation \eqref{DWT2} reduces to the Hasegawa-Mima equation \citep{HM} 
when the magnetic field is straight, $\bol{B}=B_0\bol{\nabla} z$ with $B_0\in\mathbb{R}$. 
In the remainder of this paper, we will restrict our attention to the constant density case $A_e\in\mathbb{R}$. 
To elucidate how a curved inhomogeneous magnetic field modifies the Hasegawa-Mima equation, it is convenient to study equation \eqref{DWT2} in the limit of a magnetic field of the type
\begin{equation}
\bol{B}=B_0\bol{\nabla} z+\epsilon\bol{a},\label{B1}
\end{equation}
where $\epsilon<<1$ is an ordering parameter and $\bol{a}=\lr{a_x,a_y,a_z}$ a vector field such that $\bol{\nabla}\cdot\bol{a}=0$. 
For simplicity, we assume that $a_z=\bol{a}\cdot\bol{\nabla} z=0$.
Notice that the magnetic field \eqref{B1} satisfies \eqref{orX1} and thus \eqref{orX} as well. 
Furthermore, at the first order in $\epsilon$
the curvature of the magnetic field \eqref{B1} is given by
\begin{equation}
\bol{\kappa}=\frac{\bol{B}}{B}\cdot\bol{\nabla}\lr{\frac{\bol{B}}{B}}=\frac{\epsilon}{B_0}\frac{\p\bol{a}}{\p z}.\label{curv}
\end{equation}
It is convenient to introduce the $2$-dimensional gradient and Laplacian operators
\begin{equation}
\bol{\nabla}_{\lr{x,y}}f=\frac{\p f}{\p x}\bol{\nabla} x+\frac{\p f}{\p y}\bol{\nabla} y,~~~~\Delta_{\lr{x,y}}f=
\frac{\p^2 f}{\p x^2}+\frac{\p^2 f}{\p y^2},
\end{equation}
where $f=f\lr{x,y,z}$ is some function.  
Then, 
at the first order in $\epsilon$ we have
\begin{subequations}
\begin{align}
B^2=&B_0^2,\\
\bol{\nabla}_\perp\phi=&\bol{\nabla}_{\lr{x,y}}\phi-\frac{\epsilon}{B_0}\left[
\lr{\bol{a}\cdot\bol{\nabla}\phi}\bol{\nabla} z+\frac{\p\phi}{\p z}\bol{a}
\right],\\
\bol{v}_{\bol{E}}=&\frac{\bol{\nabla} z\cp\bol{\nabla}\phi}{B_0}+\frac{\epsilon}{B_0^2}\bol{a}\cp\bol{\nabla}\phi,\\
\bol{\nabla}\cdot\bol{v}_{\bol{E}}=&\frac{\epsilon}{B_0^2}\bol{\nabla}\phi\cdot\bol{\nabla}\cp\bol{a},\\
\bol{\nabla}\cp\bol{v}_{\bol{E}}=&\frac{\Delta_{\lr{x,y}}\phi}{B_0}\bol{\nabla} z
+\frac{\epsilon}{B_0^2}\bol{\nabla}\cp\lr{\bol{a}\cp\bol{\nabla}\phi}-\frac{1}{B_0}\bol{\nabla}_{\lr{x,y}}\frac{\p \phi}{\p z}.
\end{align}
\end{subequations}
Hence, defining the bracket
\begin{equation}
\left[f,g\right]_{\lr{x,y}}=\frac{\p f}{\p x}\frac{\p g}{\p y}-\frac{\p f}{\p y}\frac{\p g}{\p x},
\end{equation}
with $f,g$ some functions, 
equation \eqref{DWT2} becomes
\begin{equation}
\begin{split}
\frac{\p}{\p t}&\left[
\lambda\phi-\frac{\sigma}{B_0^2}\Delta_{\lr{x,y}}\phi
+\frac{\epsilon\sigma}{B_0^3}\frac{\p\bol{a}}{\p z}\cdot\bol{\nabla}\phi+\frac{2\epsilon\sigma}{B_0^3}\bol{a}\cdot\bol{\nabla}\lr{\frac{\p\phi}{\p z}}
\right]=\\&\frac{\sigma}{B_0^3}\left[
\phi,\Delta_{\lr{x,y}}\phi-\frac{2\epsilon}{B_0}\bol{a}\cdot\bol{\nabla}\lr{\frac{\p\phi}{\p z}}
\right]_{\lr{x,y}}\\&-\frac{\epsilon}{B_0^2}\lr{1-\frac{\sigma}{B_0^2}\Delta_{\lr{x,y}}\phi}\bol{\nabla}\phi\cdot\bol{\nabla}\cp\bol{a}\\&+\frac{\sigma\epsilon}{B_0^4}\bol{\nabla}\phi\cdot\bol{\nabla}\lr{\Delta_{\lr{x,y}}\phi}\cp\bol{a}.\label{DWT3}
\end{split}
\end{equation}
Recalling equation \eqref{curv}, we see that the third term on the left-hand side of \eqref{DWT3} arises from the curvature of the magnetic field. Terms involving $\frac{\p\phi}{\p z}$ describe the effect of the inhomogeneity of the electric potential along the vertical axis. The term including  $\bol{\nabla}\cp\bol{B}=\epsilon\bol{\nabla}\cp\bol{a}$ can be ascribed to the presence of 
electric current in the system. 
Finally, the last term on the right-hand side results from the polarization drift associated with the component of the magnetic field $\epsilon\bol{a}$. 
Observe that equation \eqref{DWT3} reduces to the Hasegawa-Mima equation when $\epsilon=0$,
\begin{equation}
\frac{\p}{\p t}\lr{\lambda\phi-\frac{\sigma}{B_0^2}\Delta_{\lr{x,y}}\phi}=\frac{\sigma}{B_0^3}\left[\phi,\Delta_{\lr{x,y}}\phi\right]_{\lr{x,y}}.\label{HM}
\end{equation}

The effect of the field curvature \eqref{curv} can be made explicit by expanding $\bol{a}$ in Taylor series around $z=0$, and by considering the dynamics on such plane. at the first order in $z$, one has
\begin{equation}
\bol{a}=\bol{a}_0+z\bol{a}_1,~~~~\bol{\kappa}=\frac{\epsilon}{B_0}\bol{a}_1,\label{a1k1}
\end{equation}
where $\bol{a}_0=\lr{a_{0x},a_{0y},0}$ and $\bol{a}_{1}=\lr{a_{1x},a_{1x},0}$ are vector fields independent of $z$. 
Using \eqref{a1k1} and setting $\frac{\p\phi}{\p z}=0$, all terms in \eqref{DWT3} lose the dependence on $z$, giving a $2$-dimensional equation
\begin{equation}
\begin{split}
\frac{\p}{\p t}\left[
\lambda\phi-\frac{\sigma}{B_0^2}\Delta_{\lr{x,y}}\phi+\frac{\sigma}{B_0^2}\bol{\kappa}\cdot\bol{\nabla}\phi
\right]=&\frac{\sigma}{B_0^3}\left[
\phi,\Delta_{\lr{x,y}}\phi
\right]_{\lr{x,y}}\\&-\frac{1}{B_0}\lr{1-\frac{\sigma}{B_0^2}\Delta_{\lr{x,y}}\phi}\bol{\kappa}\cp\bol{\nabla}\phi\cdot\bol{\nabla} z.\label{DWTk}
\end{split}
\end{equation}

\begin{figure}
\hspace*{-0cm}\centering
\includegraphics[scale=0.3]{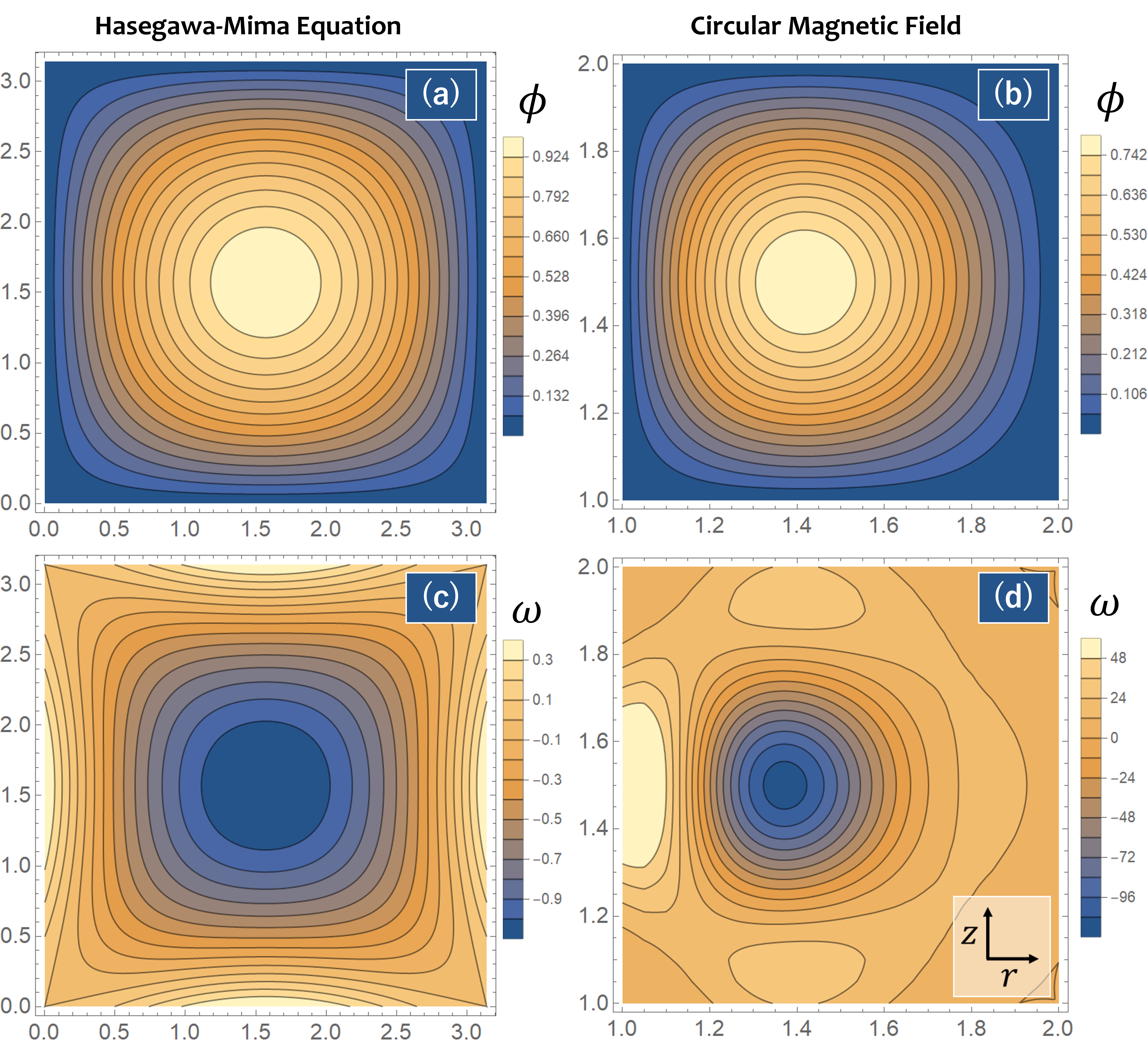}
\caption{\footnotesize (a) and (b): Contour plot of the self-organized steady electric potential $\phi$ in a straight 
magnetic field $\bol{B}=B_0\bol{\nabla} z$ (Hasegawa-Mima equation) and in a circular magnetic field $\bol{B}= B_0 r\bol{\nabla}\varphi$. 
(c) and (d): Contour plot of the respective vorticities $\omega=\bol{\nabla}\cp\bol{v}\cdot\bol{\nabla} z$ and $\omega=\bol{\nabla}\cp\bol{v}\cdot r\bol{\nabla}\varphi$ 
with the ion fluid velocity $\bol{v}$ defined by equation \eqref{totv2}. In this simulation $B_0=1$ and $f_0=2$ for the Hasegawa-Mima case, while $ B_0=1$, $\sigma=0.1$, and $f_0=1$ for the circular magnetic field case.}
\label{fig1}
\end{figure}

\begin{figure}
\hspace*{-0cm}\centering
\includegraphics[scale=0.3]{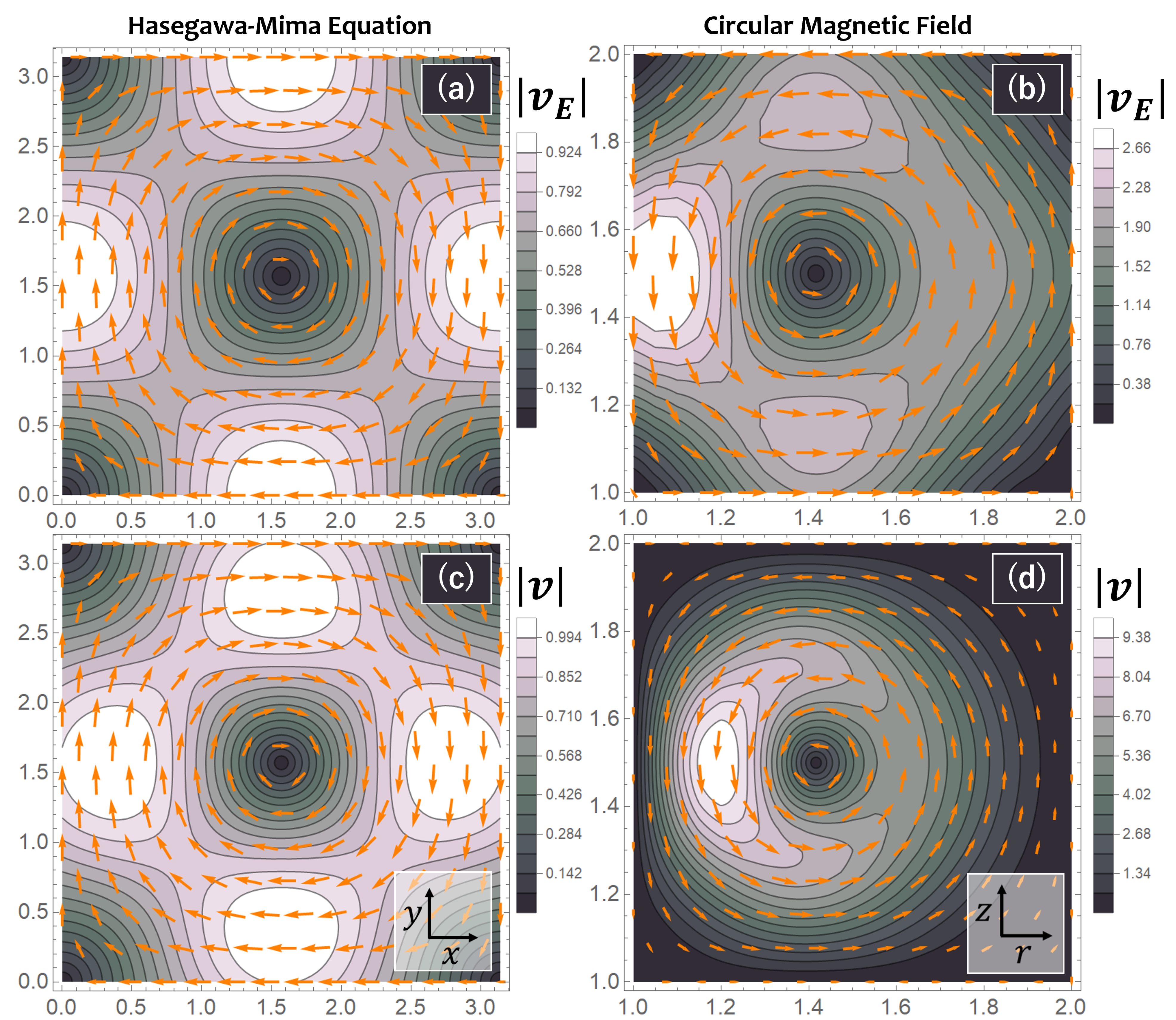}
\caption{\footnotesize (a) and (b): Vector plot of the $\bol{E}\cp\bol{B}$ drift velocity $\bol{v}_{\bol{E}}$ associated with the steady states 
of figure \ref{fig1} and contour plot of the modulus $\abs{\bol{v}_{\bol{E}}}$. (c) and (d): Vector plot of the corresponding total ion fluid velocity $\bol{v}$ of \eqref{totv2} and contour plot of the modulus $\abs{\bol{v}}$. 
}
\label{fig2}
\end{figure}

In many applications, the curvature of the magnetic field is not small. 
It is therefore instructive to examine how the Hasegawa-Mima equation is modified  
when curvature is a leading order term. To this end, consider 
a circular magnetic field $\bol{B}= B_0 r\bol{\nabla}\varphi$, 
where $\lr{r,\varphi,z}$ are cylindrical coordinates and $ B_0\in\mathbb{R}$. 
Notice that $\bol{B}^2= B_0^2$ is constant, and that the curvature of the magnetic field is given by $\bol{\kappa}=-\bol{\nabla}\log r$. 
Assuming that the condition \eqref{orX}, which in this case can be explicitly written as $r^{-1}\p\abs{\bol{\nabla}_{\perp}\phi}^2/\p z\sim\epsilon^3$, is initially satisfied by the electric potential $\phi$, 
equation \eqref{DWT2} can be written as 
\begin{equation}
\frac{\p}{\p t}\left[\lambda\phi-\frac{\sigma}{ B_0^2}\Delta_{\lr{z,r}}\phi\right]=
\frac{\kappa}{ B_0}\frac{\p\phi}{\p z}\lr{\frac{\sigma}{ B_0^2}\Delta_{\lr{z,r}}\phi-1}-\frac{\sigma\kappa}{ B_0^3 }\left[\phi,\frac{\p\phi}{\p r}\right]_{\lr{z,r}}+\frac{\sigma}{ B_0^3}\left[\phi,\Delta_{\lr{z,r}}\phi\right]_{\lr{z,r}},\label{HMcirc}
\end{equation}
where we introduced the differential operators 
\begin{equation}
\bol{\nabla}_{\lr{z,r}}f=\frac{\p f}{\p z}\bol{\nabla} z+\frac{\p f}{\p r}\bol{\nabla} r,~~~~\Delta_{\lr{z,r}}f=\frac{1}{r}\frac{\p}{\p r}\lr{r\frac{\p f}{\p r}}+\frac{\p^2 f}{\p z^2},~~~~\left[f,g\right]_{\lr{z,r}}=\frac{\p f}{\p z}\frac{\p g}{\p r}-\frac{\p g}{\p z}\frac{\p f}{\p r}.
\end{equation}
Notice that the field curvature modifies the Hasegawa-Mima equation through the modulus
$\kappa^2=1/r^2$, and that equation \eqref{HMcirc} is $2$-dimensional if axial symmetry $\p\phi/\p\varphi=0$ is assumed.
Furthermore, since we expect the electric potential $\phi$, the electric field $-\bol{\nabla}_{\lr{z,r}}\phi$, and the electric charge $-\Delta_{\lr{z,r}}\phi$ to be bounded, the Hasegawa-Mima equation can be recovered in the limit $r\rightarrow\infty$ ($\kappa\rightarrow 0$) where field lines become progressively straight.
Next, observe that in the Hasegawa-Mima equation \eqref{HM} steady states are given by the equation $\left[\phi,\Delta_{\lr{x,y}}\phi\right]=0$, or simply $\Delta_{\lr{x,y}}\phi=-f\lr{\phi}$ with $f$ some function of $\phi$. Since the electric charge $-\Delta_{\lr{x,y}}\phi$ should vanish when $\phi=0$, 
for small $\phi$ we may set  $f=f_0\phi$ with $f_0$ a positive  real constant (the positive sign physically means that charge density gradients $-\bol{\nabla}_{\lr{x,y}}\Delta_{\lr{x,y}}\phi$ are directed in the opposite direction to the electric field $-\bol{\nabla}_{\lr{x,y}}\phi$). Then, considering a $2$-dimensional domain $\lr{x,y}\in V=\left[0,\pi\right]^2$ such that the electric potential is grounded on the boundary, i.e. $\phi=0$ on $\p V$, solution of the steady Hasegawa-Mima equation gives self-organized states of the type $\phi=\phi_0\sin\lr{m x}\sin\lr{n y}$ with $\phi_0\in\mathbb{R}$, $m,n\in\mathbb{Z}$, and $f_0=m^2+n^2$. Since $f_0$ corresponds to the ratio between enstrophy and energy in the fluid limit $\lambda=0$, at equilibrium the minimum possible value $f_0=2$ corresponding to $m=n=1$ is expected to be preferentially selected (see e.g. \citep{Hasegawa85} on this point).   
Let us see how the situation changes in a curved magnetic field. Since the right-hand side of \eqref{HMcirc} can be written as
\begin{equation}
\frac{1}{r}\left[\phi,-\frac{r}{ B_0}-\frac{\sigma}{ B_0^3}\frac{\p\phi}{\p r}+\frac{\sigma }{ B_0^3}r\Delta_{\lr{z,r}}\phi\right]_{\lr{z,r}}, 
\end{equation}
steady states in the circular magnetic field $\bol{B}= B_0 r\bol{\nabla}\phi$ are described by the equation 
\begin{equation}
\frac{\sigma}{ B_0^3}\lr{\frac{\p^2\phi}{\p r^2}+\frac{\p^2\phi}{\p z^2}}=\frac{1}{ B_0}-\frac{f\lr{\phi}}{r}.\label{HMcircEq}
\end{equation}
Considering the case $f=f_0\phi$ and taking a $2$-dimensional toroidal domain with squared cross section  $\lr{r,z}\in V=\left[1,2\right]^2$ and Dirichlet boundary conditions $\phi=0$ on the boundary $\p V$, solution of \eqref{HMcircEq} results in a self-organized steady state sustained by the
curvature of the magnetic field. Figure \ref{fig1} shows contour plots of the 
steady electric potential $\phi$ and vorticity $\omega=\bol{\nabla}\cp\bol{v}\cdot r\bol{\nabla}\varphi$ with $\bol{v}$ given by \eqref{totv} obtained by numerical solution of \eqref{HMcircEq} as compared with the Hasegawa-Mima case. The corresponding
fluid drifts $\bol{v}_{\bol{E}}$ and $\bol{v}$ are given in figure \ref{fig2}. 
From these figures, one sees that contours of electric potential $\phi$, vorticity $\omega$, and fluid drifts $\abs{\bol{v}_{\bol{E}}}$ and $\abs{\bol{v}}$ tend to accumulate in regions of higher curvature (small radius $r$), implying the onset of steeper gradients, 
while inhomogeneities are suppressed where the curvature is weaker. Figure \ref{fig2} also reveals that the fluid drifts $\bol{v}_{\bol{E}}$ and $\bol{v}$ are enhanced 
where the curvature is stronger. 
We remark that the observed behavior is not caused by the gradient or curvature drifts of the guiding center framework because the cold ions approximation is being considered. In addition, the background magnetic field has uniform strength in this example. Hence, curvature 
affects self-organized states through the $\bol{E}\times\bol{B}$ drift, as clear from equation \eqref{DWT2}. 

\begin{figure}
\hspace*{-0cm}\centering
\includegraphics[scale=0.3]{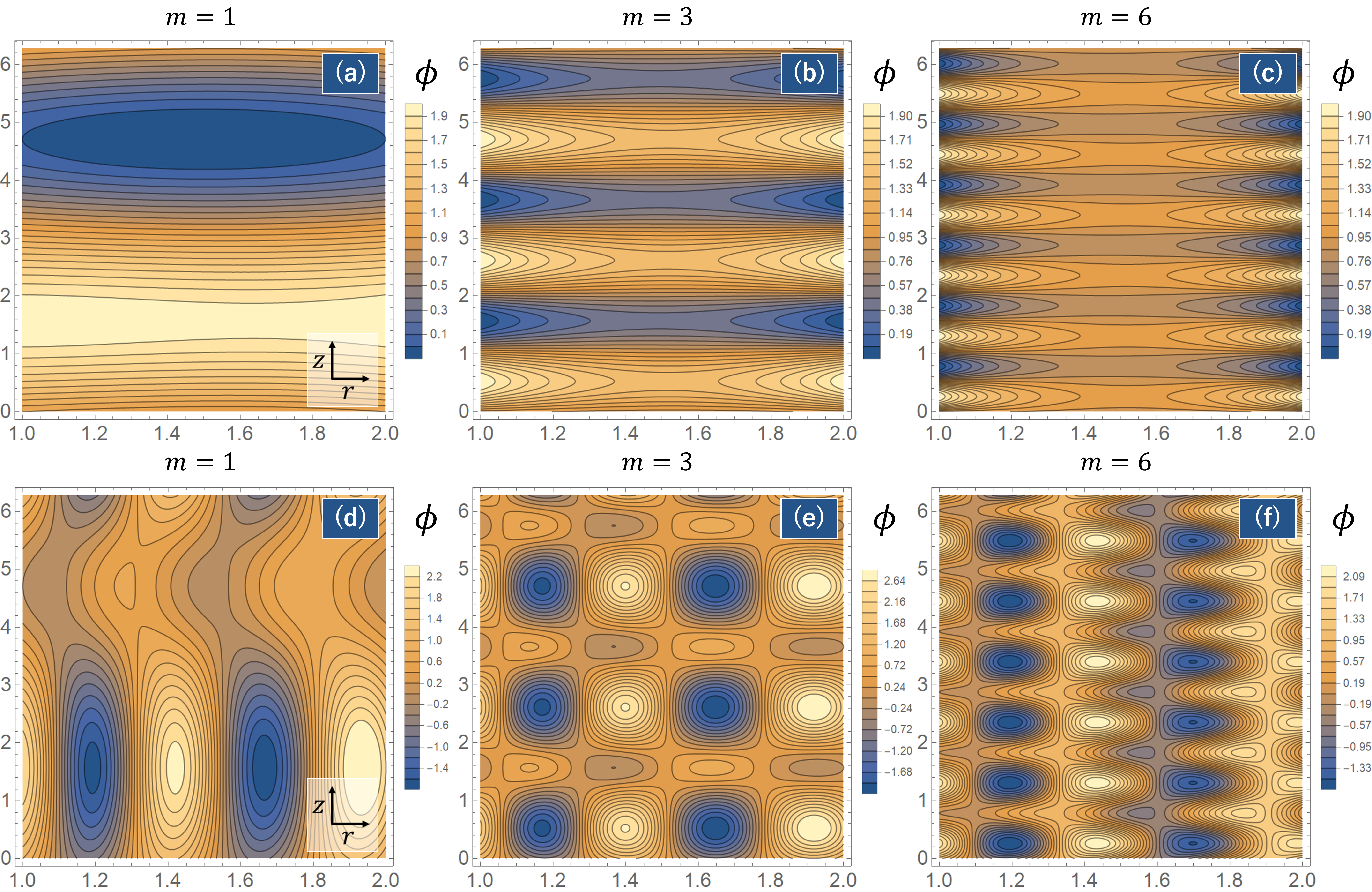}
\caption{\footnotesize Contour plots of the electric potential $\phi$ obtained by solution of \eqref{HMcircz} for different values of $m$ and $ B_0$. Dirichlet boundary conditions $p(1)=p(2)=1$ and $q(1)=q(2)=1$ are used. (a) and (d): The case $m=1$ for $ B_0=1$ (above) and $ B_0=5$ (below). (b) and (e): The case $m=3$ for $ B_0=1$ (above) and $ B_0=5$ (below). (c) and (f): The case $m=6$ for $ B_0=1$ (above) and $ B_0=5$ (below). In this simulation $\sigma=0.5$ and $f_0=1$ .}
\label{fig3}
\end{figure}

The effect of vertical inhomogeneity in the potential $\phi$ can be further examined by setting $\phi=q+\sin\lr{m z}p$, with $q\lr{r}$ and $p\lr{r}$ radial functions and $m\in\mathbb{Z}$. Then, equation \eqref{HMcircEq} reduces to the system
\begin{equation}
\frac{\sigma}{ B_0^3}\frac{\p^2 p}{\p r^2}+\lr{\frac{f_0}{r}-m^2\frac{\sigma}{ B_0^3}}p=0,~~~~\frac{\sigma}{ B_0^3}\frac{\p^2 q}{\p r^2}+\frac{f_0}{r}q-\frac{1}{ B_0}=0.\label{HMcircz}
\end{equation}
A plot of the electric potential $\phi$ 
obtained by solution of system \eqref{HMcircz} 
for different values of $m$ and $ B_0$ is given in figure \ref{fig3}.
Notice that for a sufficiently large magnetic field strength $ B_0$ structures arise in the radial direction as well. The size and spacing of these radial structures appears to be modulated by the background strength and  curvature of the magnetic field rather than the constant $m$ associated with the vertical oscillation.

\section{The case of curved magnetic fields crossing a surface perpendicularly}
The Hasegawa-Mima equation is endowed with two inviscid invariants, the total energy and the generalized enstrophy. 
The generalized enstrophy is an invariant arising from the $2$-dimensional nature of the governing equation, 
which is restricted to the flat $\lr{x,y}$ plane with normal given by the straight vertical magnetic field $\bol{B}=B_0\bol{\nabla} z$. 
It is useful to consider the conditions under which the same kind of topological invariant persists in general magnetic fields. 
A necessary condition for the analogy with $2$-dimensional vorticity dynamics to apply 
is that the magnetic field defines the normal direction of a general (not necessarily flat) $2$-dimensional surface $\Sigma\subset\mathbb{R}^3$. 
In this case, the dynamics is restricted to the surface $\Sigma$ because the drift velocity \eqref{totv} satisfies $\bol{B}\cdot\bol{v}=0$.
The geometric condition for a vector field $\bol{B}$ to locally define the normal of a surface $\Sigma$ is given by the Frobenius integrability condition \citep{Frankel},
\begin{equation}
\bol{B}\cdot\bol{\nabla}\cp\bol{B}=0.\label{hel}
\end{equation}
In particular, if the magnetic field $\bol{B}$ has vanishing helicity density then there exist locally defined functions $\alpha,C$ such that
\begin{equation}
\bol{B}=\alpha\bol{\nabla} C.\label{IntB}
\end{equation}
The magnetic field $\bol{B}$ thus defines the normal to the surface $C={\rm constant}$. 
When $\alpha=\alpha\lr{C}$, the magnetic field $\bol{B}$ becomes a vacuum magnetic field because $\bol{\nabla}\cp\bol{B}=\bol{0}$.  
In this section, we will be concerned with magnetic fields of the type \eqref{IntB}. We will assume that the functions $\alpha$ and $C$ exist in the domain $ V$, and that $\bol{B}\neq\bol{0}$ in $ V$. 
Next, recall that the magnetic field $\bol{B}$ must be solenoidal. 
Hence, the Lie-Darboux theorem \citep{deLeon,Arnold} applies: locally there exist functions $\Psi,\theta$ such that
\begin{equation}
\bol{B}=\bol{\nabla}\Psi\cp\bol{\nabla}\theta.
\end{equation}
Again, we will assume that the functions $\Psi$ and $\theta$ are well defined in the domain $ V$. 
A typical example of vacuum magnetic field is the magnetic field generated by a point dipole. In this case 
\begin{equation}
C=-M\frac{z}{\lr{r^2+z^2}^{3/2}},~~~~\Psi=M\frac{r^2}{\lr{r^2+z^2}^{3/2}},~~~~\theta=\varphi,
\end{equation}
where $\lr{r,\varphi,z}$ are cylindrical coordinates and $M$ a physical constant with units of $Tm^3$.
The functions $\lr{C,\Psi,\theta}$ can be used as a system of curvilinear coordinates. The Jacobian determinant of the coordinate transformation is given by
\begin{equation}
J=\bol{\nabla} C\cdot\bol{\nabla}\Psi\cp\bol{\nabla}\theta=\frac{B^2}{\alpha}.
\end{equation}
Given two functions $f,g$ it is convenient to introduce the bracket
\begin{equation}
\left[f,g\right]_{\lr{\Psi,\theta}}=\frac{\p f}{\p\Psi}\frac{\p g}{\p\theta}-
\frac{\p f}{\p\theta}\frac{\p g}{\p\Psi}.\label{bra2}
\end{equation}
Using \eqref{bra2}, the derived equation \eqref{DWT2} can be written as
\begin{equation}
\begin{split}
\frac{\p}{\p t}\left[\lambda\phi-\sigma\bol{\nabla}\cdot\lr{\frac{\bol{\nabla}_{\perp}\phi}{B^2}}\right]=&\frac{B^2}{\alpha}\left[\phi,
-\frac{\alpha}{B^2}
+\sigma\frac{\alpha^2}{B^4}\bol{\nabla}\cdot\lr{\frac{\bol{\nabla}_{\perp}\phi}{\alpha}}
\right]_{\lr{\Psi,\theta}}.
\label{DWT4}
\end{split}
\end{equation}
One can verify that under appropriate boundary conditions on the surface boundary $\p\Sigma$ the surface energy
\begin{equation}
H_{\Sigma}=\frac{1}{2}\int_{\Sigma}\lr{\lambda\phi^2+\sigma\frac{\abs{\bol{\nabla}_{\perp}\phi}^2}{B^2}}\frac{\alpha}{B^2}\,{\rm d}\Psi {\rm d}\theta,\label{H2}
\end{equation}
is a constant of \eqref{DWT4}.
The condition for conservation of generalized enstrophy,
\begin{equation}
W_{\Sigma}=\frac{1}{2}\int_{\Sigma}\left\{
\lambda\frac{\abs{\bol{\nabla}_{\perp}\phi}^2}{B^2}+\sigma\left[\bol{\nabla}\cdot\lr{\frac{\bol{\nabla}_{\perp}\phi}{B^2}}\right]^2
\right\}\frac{\alpha}{B^2}\,{\rm d}\Psi {\rm d}\theta,\label{W}
\end{equation}
can be obtained by noting that the second argument of the bracket on the right-hand side of equation \eqref{DWT4} 
must be a function of $\bol{\nabla}\cdot\lr{B^{-2}\bol{\nabla}_{\perp}\phi}$ 
up to a function of $\phi$. The condition is: 
\begin{equation}
\alpha=\frac{k}{\abs{\bol{\nabla} C}^2}=\frac{B^2}{k},~~~~k\in\mathbb{R},\label{conda}
\end{equation}
which is equivalent to
\begin{equation}
\bol{\nabla}\times\lr{\frac{\bol{B}}{B^2}}=\bol{0}.
\end{equation}
In this case, $\bol{\nabla}\cdot\bol{v}_{\bol{E}}=\bol{\nabla}\phi\cdot\bol{\nabla}\cp\lr{B^{-2}\bol{B}}=0$. 
Notice also that $\bol{\nabla}\cdot\bol{B}=0$ implies that configurations of the type \eqref{conda} must satisfy
\begin{equation}
\bol{\nabla}\cdot\lr{\frac{\bol{\nabla} C}{\abs{\bol{\nabla} C}^2}}=0.\label{condB}
\end{equation}
In spherical geometry, denoting with $R$ the spherical radius, a magnetic field satisfying \eqref{conda} and \eqref{condB} can be obtained by setting 
\begin{equation}
\alpha=\frac{ B_0}{R^4},~~~~ B_0\in\mathbb{R},~~~~C=\frac{R^3}{3}.\label{sp}
\end{equation}
Similarly, in cylindrical geometry magnetic fields compatible with \eqref{conda} and \eqref{condB} include those generated by 
\begin{subequations}
\begin{align}
\alpha&=\frac{ B_0}{r^2},~~~~ B_0\in\mathbb{R},~~~~C=\frac{r^2}{2},\label{cyl}\\
\alpha&= B_0 r^2,~~~~ B_0\in\mathbb{R},~~~~C=\varphi.\label{cyl2}
\end{align}
\end{subequations}
Observe that when $\lambda=0$ equation \eqref{DWT4} with either \eqref{sp} or \eqref{cyl} 
gives the usual $2$-dimensional vorticity dynamics on a sphere or cylinder respectively.   
The case of equation \eqref{cyl2} corresponds to a circular magnetic field $\bol{B}= B_0 r^2\bol{\nabla}\varphi$ 
with curvature $\bol{\kappa}=-\bol{\nabla}\log r$. Assuming axial symmetry $\phi=\phi\lr{r,z}$, 
the corresponding form of equation \eqref{DWT2} is again $2$-dimensional, 
\begin{equation}
\frac{\p}{\p t}\left[\lambda\phi-\frac{\sigma}{ B_0^2}\bol{\nabla}\cdot\lr{\kappa^2\bol{\nabla}_{\lr{z,r}}\phi}\right]=
\frac{\sigma}{ B_0^3}\kappa\left[\phi,\bol{\nabla}\cdot\lr{\kappa^2\bol{\nabla}_{\lr{z,r}}\phi}\right]_{\lr{z,r}}.
\end{equation}

\section{Concluding remarks}
In conclusion, we have derived a model equation \eqref{DWT2} describing electrostatic plasma turbulence in a general magnetic field.
The equation preserves the mass \eqref{MV} and the energy \eqref{HV}, and reduces to the Hasegawa-Mima equation in the limit of a straight magnetic field. 
The ordering adopted in the derivation of the equation is analogous to the classical one, 
exception made for two hypothesis. 
On one hand, a weaker condition is assumed on the time scale, equation \eqref{tv1v2}, which replaces 
the usual scaling $\omega_d/\Omega_c\sim\epsilon$ to allow first order contributions in $\phi_t$ originating from the non-vanishing of $\bol{\nabla}\cdot\bol{v}_{\bol{E}}$ in a general magnetic field. On the other hand,   
the requirement \eqref{orX}, which is automatically satisfied in the standard setting of the Hasegawa-Mima equation, ensures the consistency of the ordering with respect to preservation of energy. 
This latter condition implies that the derived equation is best suited for magnetic fields satisfying \eqref{orX1}, implying a second order $\bol{E}\times\bol{B}$ drift velocity divergence, or
for spatial scales such that the change in $\bol{v}_{\bol{E}}^2$ represents a third order contribution \eqref{orX2}. 

Conservation of generalized enstrophy
holds when the magnetic field defines the normal of a surface and it is compatible with $2$-dimensional vorticity dynamics. 
It should be noted that inverse energy cascades are expected to occur for all magnetic configurations such that generalized enstrophy is constant. Indeed, the order of $\phi$ derivatives appearing in \eqref{H2} and \eqref{W} is not changed by the geometry of the background magnetic field. 
Detailed analysis of the effect of magnetic topology on inverse cascades and zonal flows is left for future work. 


As a physical application of the obtained equation, we have studied how the curvature of a circular magnetic field with uniform strength modifies self-organized steady states 
by comparison with analogous equilibria in a straight magnetic field. A strong curvature tends to attract level sets of electric potential and vorticity, and to enhance fluid drift velocity. A higher curvature also  
appears to contribute to higher heterogeneity in the electric potential. 
This behavior cannot be ascribed to gradient and curvature drifts occuring in the guiding center picture 
because the model relies on the cold ions approximation. Hence, the effect of curvature is mediated by the $\bol{E}\cp\bol{B}$ drift.

Finally, the results reported in this paper
may be useful to construct simplified models of turbulence in complex plasma systems of practical interest. In particular, we expect the derived equation to be appropriate for the description of turbulent states in the core of $H$-mode plasmas where density frequently exhibits a flat profile, and more generally of turbulence in those magnetic confinement systems whose macroscopic density may be well approximated by a steady spatial function $A_e\lr{\bol{x}}$ and a locally thermalized electron distribution.  



\section*{Acknowledgment}
The research of NS was partially supported by JSPS KAKENHI Grant No. 21K13851 and No. 17H01177.

\bibliographystyle{jpp}
\bibliography{jpp-instructions}

\begin{thebibliography}{42}
\expandafter\ifx\csname natexlab\endcsname\relax\def\natexlab#1{#1}\fi
\def\au#1{#1} \def\ed#1{#1} \def\yr#1{#1}\def\at#1{#1}\def\jt#1{\textit{#1}}
  \def\bt#1{#1}\def\bvol#1{\textbf{#1}} \def\vol#1{#1} \def\pg#1{#1}
  \def\publ#1{#1}\def\arxiv#1{#1}\def\org#1{#1}\def\st#1{\textit{#1}}

\bibitem[Arnold(1989)]{Arnold}
{\sc \au{Arnold, V.~I.}} \at{ \yr{1989} } \bt{In {\em Mathematical methods of
  classical mechanics\/}},  \pg{pp. 230--232}.  \publ{Springer}.

\bibitem[Batchelor(1969)]{Batchelor}
{\sc \au{Batchelor, G.~K.}} \yr{1969}  \at{Computation of the energy spectrum
  in homogeneous two-dimensional turbulence}.  \jt{The Physics of Fluids}
  \bvol{12}~(II),  \pg{233--239}.

\bibitem[Bernert {\em et~al.\/}(2015)Bernert, Eich, Kallenbach, Carralero,
  Huber, Lang, Potzel, Reimold, Schweinzer, Viezzer, Zohm \& the ASDEX
  Upgrade~team]{Bernert}
{\sc \au{Bernert, M.}, \au{Eich, T.}, \au{Kallenbach, A.}, \au{Carralero, D.},
  \au{Huber, A.}, \au{Lang, P.~T.}, \au{Potzel, S.}, \au{Reimold, F.},
  \au{Schweinzer, J.}, \au{Viezzer, E.}, \au{Zohm, H.} \& \au{the ASDEX
  Upgrade~team}} \yr{2015}  \at{The h-mode density limit in the full tungsten
  asdex upgrade tokamak}.  \jt{Plasma Phys. Control. Fusion}  \bvol{57},
  \pg{1--12}.

\bibitem[Brizard \& Hahm(2007)]{Brizard}
{\sc \au{Brizard, A.~J.} \& \au{Hahm, T.~S.}} \yr{2007}  \at{Foundations of
  nonlinear gyrokinetic theory}.  \jt{Rev. Mod. Phys.}  \bvol{79},
  \pg{421--468}.

\bibitem[Cary \& Brizard(2009)]{Cary}
{\sc \au{Cary, J.~R.} \& \au{Brizard, A.~J.}} \yr{2009}  \at{Hamiltonian theory
  of guiding-center motion}.  \jt{Rev. Mod. Phys.}  \bvol{81}~(2),
  \pg{693--738}.

\bibitem[Charney(1948)]{Charney}
{\sc \au{Charney, J.~G.}} \yr{1948}  \at{On the scale of atmospheric motions}.
  \jt{Geof.~Publ.}  \bvol{17}~(2),  \pg{3--17}.

\bibitem[Charney \& Drazin(1961)]{Charney3}
{\sc \au{Charney, J.~G.} \& \au{Drazin, P.~G.}} \yr{1961}  \at{Propagation of
  planetary-scale disturbances from the lower into the upper atmosphere}.
  \jt{J.~Geophys.~Res.}  \bvol{66}~(1),  \pg{83--109}.

\bibitem[Diamond {\em et~al.\/}(2011)Diamond, Hasegawa \& Mima]{Diamond}
{\sc \au{Diamond, P.~H.}, \au{Hasegawa, A.} \& \au{Mima, K.}} \yr{2011}
  \at{Vorticity dynamics, drift wave turbulence, and zonal flows: a look back
  and a look ahead}.  \jt{Plasma Phys. Control. Fusion}  \bvol{53}~(124001),
  \pg{1--23}.

\bibitem[Dritschel {\em et~al.\/}(2015)Dritschel, Qi \& Marston]{Dritschel}
{\sc \au{Dritschel, D.~G.}, \au{Qi, W.} \& \au{Marston, J.~B.}} \yr{2015}
  \at{On the late-time behavior of a bounded, inviscid two-dimensional flow}.
  \jt{J.~Fluid Mech.}  \bvol{783},  \pg{1--22}.

\bibitem[Dubin {\em et~al.\/}(1983)Dubin, Krommes \& Oberman]{Dubin}
{\sc \au{Dubin, D. H.~E.}, \au{Krommes, J.~A.} \& \au{Oberman, C.}} \yr{1983}
  \at{Nonlinear gyrokinetic equations}.  \jt{The Physics of Fluids}  \bvol{26},
   \pg{3524--3535}.

\bibitem[Frankel(2012)]{Frankel}
{\sc \au{Frankel, T.}} \at{ \yr{2012} } \bt{In {\em The Geometry of
  Physics\/}},  \pg{pp. 165--171}.  \publ{Cambridge University Press}.

\bibitem[Fujisawa(2004)]{Fujisawa}
{\sc \au{Fujisawa, A. et~al.}} \yr{2004}  \at{Identification of zonal flows in
  a toroidal plasma}.  \jt{Phys. Rev. Lett.}  \bvol{93}~(165002),  \pg{1--4}.

\bibitem[Hahm(1996)]{Hahm962}
{\sc \au{Hahm, T.~S.}} \yr{1996}  \at{Nonlinear gyrokinetic equations for
  turbulence in core transport barriers}.  \jt{Physics of Plasmas}  \bvol{3},
  \pg{4658--4664}.

\bibitem[Hahm \& Tang(1996)]{Hahm96}
{\sc \au{Hahm, T.~S.} \& \au{Tang, W.~M.}} \yr{1996}  \at{Nonlinear theory of
  collisionless trapped ion modes}.  \jt{Physics of Plasmas}  \bvol{3},
  \pg{242--247}.

\bibitem[Hahm {\em et~al.\/}(2009)Hahm, Wang \& Madsen]{Hahm09}
{\sc \au{Hahm, T.~S.}, \au{Wang, L.} \& \au{Madsen, J.}} \yr{2009}  \at{Fully
  electromagnetic nonlinear gyrokinetic equations for tokamak edge turbulence}.
   \jt{Physics of Plasmas}  \bvol{16}~(022305),  \pg{1--11}.

\bibitem[Hasegawa(2004)]{Hasegawa85}
{\sc \au{Hasegawa, A.}} \yr{2004}  \at{Self-organization processes in
  continuous media}.  \jt{Adv. Physics}  \bvol{34}~(1),  \pg{1--42}.

\bibitem[Hasegawa \& Mima(1977{\natexlab{{\em a\/}}})]{HM}
{\sc \au{Hasegawa, A.} \& \au{Mima, K.}} \yr{1977{\natexlab{{\em a\/}}}}
  \at{Pseudo-three-dimensional turbulence in magnetized nonuniform plasma}.
  \jt{Phys.~Fluids}  \bvol{21}~(1),  \pg{87--92}.

\bibitem[Hasegawa \& Mima(1977{\natexlab{{\em b\/}}})]{HM2}
{\sc \au{Hasegawa, A.} \& \au{Mima, K.}} \yr{1977{\natexlab{{\em b\/}}}}
  \at{Stationary spectrum of strong turbulence in magnetized nonuniform
  plasma}.  \jt{Phys.~Rev.~Lett.}  \bvol{39}~(4),  \pg{205--208}.

\bibitem[Hasegawa \& Mima(2018)]{HM5}
{\sc \au{Hasegawa, A.} \& \au{Mima, K.}} \yr{2018}  \at{Strong turbulence,
  self-organization and plasma confinement}.  \jt{Eur. Phys. J. H}  \bvol{43},
  \pg{499--521}.

\bibitem[Hasegawa \& Wakatani(1983)]{HM4}
{\sc \au{Hasegawa, A.} \& \au{Wakatani, M.}} \yr{1983}  \at{Plasma edge
  turbulence}.  \jt{Phys. Rev. Lett.}  \bvol{50}~(9),  \pg{682--686}.

\bibitem[Hasegawa \& Wakatani(1987)]{HM3}
{\sc \au{Hasegawa, A.} \& \au{Wakatani, M.}} \yr{1987}  \at{Self-organization
  of electrostatic turbulence in a cylibdrical plasma}.  \jt{Phys. Rev. Lett.}
  \bvol{59}~(14),  \pg{1581--1584}.

\bibitem[Hazeltine(1983)]{Hazeltine}
{\sc \au{Hazeltine, R.~D.}} \yr{1983}  \at{Reduced magnetohydrodynamics and the
  hasegawa-mima equation}.  \jt{Phys. Fluids}  \bvol{26}~(11),
  \pg{3242--3245}.

\bibitem[Hazeltine {\em et~al.\/}(1987)Hazeltine, Hsu \& Morrison]{Hazeltine2}
{\sc \au{Hazeltine, R.~D.}, \au{Hsu, C.~T.} \& \au{Morrison, P.~J.}} \yr{1987}
  \at{Hamiltonian four-field model for nonlinear tokamak dynamics}.  \jt{The
  Physics of Fluids}  \bvol{30}~(10),  \pg{3204--3211}.

\bibitem[Hazeltine \& Waelbroeck(1998)]{HazeltineBook}
{\sc \au{Hazeltine, R.~D.} \& \au{Waelbroeck, F.~L.}} \at{ \yr{1998} } \bt{In
  {\em The Framework of Plasma Physics\/}},  \pg{p.~67}.  \publ{Perseus Books}.

\bibitem[Horton(1999)]{Horton2}
{\sc \au{Horton, W.}} \yr{1999}  \at{Drift waves and transport}.  \jt{Rev. Mod.
  Phys.}  \bvol{71}~(3),  \pg{735--778}.

\bibitem[Horton \& Hasegawa(1994)]{Horton}
{\sc \au{Horton, W.} \& \au{Hasegawa, A.}} \yr{1994}  \at{Quasi-two-dimensional
  dynamics of plasmas and fluids}.  \jt{Chaos}  \bvol{4}~(2),  \pg{227--251}.

\bibitem[Kaufman(1986)]{Kauf}
{\sc \au{Kaufman, A.~N.}} \yr{1986}  \at{The electric dipole of a guiding
  center and the plasma momentum density}.  \jt{The Physics of Fluids}
  \bvol{29}~(5),  \pg{1736--1737}.

\bibitem[Kraichnan(1967)]{Kraichnan}
{\sc \au{Kraichnan, R.~H.}} \yr{1967}  \at{Inertial ranges in two-dimensional
  turbulence}.  \jt{The Physics of Fluids}  \bvol{10}~(7),  \pg{1417--1423}.

\bibitem[Kraichnan \& Montgomery(1980)]{Kraichnan2}
{\sc \au{Kraichnan, R.~H.} \& \au{Montgomery, D.}} \yr{1980}
  \at{Two-dimensional turbulence}.  \jt{Rep. Prog. Phys.}  \bvol{43},
  \pg{547--619}.

\bibitem[Krommes(2002)]{Krommes}
{\sc \au{Krommes, J.~A.}} \yr{2002}  \at{Fundamental statistical descriptions
  of plasma turbulence in magnetic fields}.  \jt{Phys. Rep.}  \bvol{360},
  \pg{1--352}.

\bibitem[de~L\'eon(1989)]{deLeon}
{\sc \au{de~L\'eon, M.}} \at{ \yr{1989} } \bt{In {\em Methods of differential
  geometry in analytical mechanics\/}},  \pg{pp. 250--253}.  \publ{Elsevier}.

\bibitem[Littlejohn(1981)]{Little}
{\sc \au{Littlejohn, R.~G.}} \yr{1981}  \at{Hamiltonian formulation of guiding
  center motion}.  \jt{The Physics of Fluids}  \bvol{24},  \pg{1730--1749}.

\bibitem[Morrison(1998)]{Morrison}
{\sc \au{Morrison, P.~J.}} \yr{1998}  \at{Hamiltonian description of the ideal
  fluid}.  \jt{Rev. Mod. Phys.}  \bvol{70},  \pg{467--521}.

\bibitem[Nazarenko \& Quinn(2009)]{Nazarenko}
{\sc \au{Nazarenko, S.} \& \au{Quinn, B.}} \yr{2009}  \at{Triple cascade
  behavior in quasigeostrophic and drift turbulence and generation of zonal
  jets}.  \jt{Phys. Rev. Lett.}  \bvol{103}~(118501),  \pg{1--4}.

\bibitem[Numata {\em et~al.\/}(2007)Numata, Ball \& Dewar]{Numata}
{\sc \au{Numata, R.}, \au{Ball, R.} \& \au{Dewar, R.~L.}} \yr{2007}
  \at{Bifurcation in electrostatic resistive drift wave turbulence}.  \jt{Phys.
  Plasmas}  \bvol{14}~(102312),  \pg{1--8}.

\bibitem[Rivera {\em et~al.\/}(2003)Rivera, Daniel, Chen \& Ecke]{Rivera}
{\sc \au{Rivera, M.~K.}, \au{Daniel, W.~B.}, \au{Chen, S.~Y.} \& \au{Ecke,
  R.~E.}} \yr{2003}  \at{Energy and enstrophy transfer in decaying
  two-dimensional turbulence}.  \jt{Phys. Rev. Lett.}  \bvol{90}~(10),
  \pg{1--4}.

\bibitem[Singh \& Diamond(2021)]{Singh}
{\sc \au{Singh, R.} \& \au{Diamond, P.~H.}} \yr{2021}  \at{A unified theory of
  zonal flow shears and density corrugations in drift wave turbulence}.
  \jt{Plasma Phys. Control. Fusion}  \bvol{63}~(035015),  \pg{1--28}.

\bibitem[Strinzi \& Scott(2004)]{Strinzi}
{\sc \au{Strinzi, D.} \& \au{Scott, B.}} \yr{2004}  \at{Nonlocal nonlinear
  electrostatic gyrofluid equations}.  \jt{Physics of Plasmas}  \bvol{11},
  \pg{5452--5461}.

\bibitem[Tassi {\em et~al.\/}(2009)Tassi, Chandre \& Morrison]{Tassi}
{\sc \au{Tassi, E.}, \au{Chandre, C.} \& \au{Morrison, P.~J.}} \yr{2009}
  \at{Hamiltonian derivation of the charney-hasegawa-mima equation}.  \jt{Phys.
  Plasmas}  \bvol{16}~(082301),  \pg{1--5}.

\bibitem[Wakatani \& Hasegawa(1984)]{Wakatani}
{\sc \au{Wakatani, M.} \& \au{Hasegawa, A.}} \yr{1984}  \at{A collisional drift
  wave description of plasma edge turbulence}.  \jt{The Physics of Fluids}
  \bvol{27}~(611),  \pg{611--618}.

\bibitem[Weinstein(1983)]{Weinstein}
{\sc \au{Weinstein, A.}} \yr{1983}  \at{Hamiltonian structure for drift waves
  and geostrophic flow}.  \jt{The Physics of Fluids}  \bvol{26},
  \pg{388--390}.

\bibitem[Xiao {\em et~al.\/}(2009)Xiao, Wan, Chen \& Eyink]{Xiao}
{\sc \au{Xiao, Z.}, \au{Wan, M.}, \au{Chen, S.} \& \au{Eyink, G.~L.}} \yr{2009}
   \at{Physical mechanism of the inverse energy cascade of two-dimensional
  turbulence: a numerical investigation}.  \jt{J.~Fluid Mech.}  \bvol{619},
  \pg{1--44}.

\end{thebibliography}

\end{document}